\documentclass[twocolumn, showpacs,preprintnumbers,amsmath,amssymb]{revtex4}
\usepackage{amssymb}
\usepackage{xcolor}
\usepackage{graphicx}
\usepackage{dcolumn}
\usepackage{bm}
\usepackage{multirow}
\usepackage{booktabs}
\usepackage{natbib}
\usepackage{soul}
\usepackage{amssymb}
\usepackage{amsmath}	
\usepackage{color}
\usepackage{cases} 

\usepackage{array}

\begin{document}
\preprint{APS/123-QED}

\title{Theoretical study of dissociative recombination and vibrational excitation of the BF$_2^+$ ion by an electron impact}
\author{V. Kokoouline$^{1}$}\email[]{ Viatcheslav.Kokoouline@ucf.edu}
\author{M. Ayouz$^{2}$}
\author{J. Zs Mezei$^{3,4,5}$}\email[]{mezei.zsolt@atomki.hu}
\author{K. Hassouni$^{4}$}
\author{I. F. Schneider$^{5,6}$}
\affiliation{$^{1}$Department of Physics, University of Central Florida, Orlando, FL 32816, USA}
\affiliation{$^{2}$LGPM, CentraleSupelec, Universit\'e Paris-Saclay, 91190 Gif-sur-Yvette, France}
\affiliation{$^{3}$HUN-REN Institute for Nuclear Research (ATOMKI), H-4001 Debrecen, Hungary}%
\affiliation{$^{4}$LSPM CNRS-UPR3407, Universit{\'{e}} Paris 13, 93430 Villetaneuse, France}
\affiliation{$^{5}$LOMC-UMR6294, CNRS, Universit\'e Le Havre Normandie, 76600 Le Havre, France}%
\affiliation{$^{6}$LAC-UMR9188, CNRS Universit\'e Paris-Saclay, F-91405 Orsay, France}%
\date{\today}

\begin{abstract}
Cross sections for dissociative recombination and electron-impact vibrational excitation of the BF$_2^+$ molecular ion are computed using a theoretical approach that combines the normal modes approximation for the vibrational states of the target ion and the UK R-matrix code employed to evaluate electron-ion scattering matrices for fixed geometries of the ion.
Thermally-averaged rate coefficients are obtained from the cross sections for temperatures $10-3000$~K.
\end{abstract}


\maketitle

Non-equilibrium plasmas produced with electrical discharges in BF$_3$ containing feed gas present a continuously increasing interest for a large number of applications. In particular, BF$_3$ is very often the boron carrier when plasmas are used for material processing~\cite{sennikov2017,torigoe2016,gonzatti2010}. Basically, BF$_3$ plasmas are used either for (i) the synthesis of ultra-hard boron compounds, e.g. boron carbides \cite{sennikov2017}, (ii) the deposition of boron nitride, an advanced material with a large number of functionalities \cite{torigoe2016}, and (iii) p-type doping by boron in the semi-conductor and photovoltaic industries \cite{gonzatti2010,duchaine2012}. As far as doping applications are concerned, plasma immersion ion implantation (PIII) processes are probably one of the most promising in terms of cost and technical performance \cite{duchaine2012,young2016}. These processes make use of very low pressure, very high density magnetized plasmas generated in a BF$_3$ containing feed gas. Depending on the level of the power coupled to the plasma, several cations - BF$_3^+$, BF$_2^+$, BF$^+$, B$^+$ - and anions - e.g. F$^-$ - may be produced \cite{farber1984,yongki2000}. The positive ions are then extracted from the source to an implantation chamber where the processed silicon substrate is submitted to very high negative voltage pulses. These pulses result in a large acceleration of the positive ions that are implanted in the substrate, which results in the doping of this later. The implantation depth depends on the nature and the energy distribution of the ions impinging the substrate, while the doping level depends on the ion flux and, consequently, on the plasma density. The plasma density and the relative predominance of the different ions are determined by the ionization kinetics in the source region. The plasma sources used in PIII processes are usually magnetized \cite{stewart1991}. %
The ambipolar diffusion in the radial direction that is perpendicular to the magnetic field is strongly reduced with respect to the parallel diffusion. Under typical plasmas conditions used in PIII processes, i.e., $B\approx100-500$ Gauss and $p \approx 1$ Pa, the electron cyclotron frequency is the GHz range, while the electron-neutral momentum transfer collision frequency is in the MHz range, according to Eq.~(5.4.5) of \cite{lieberman2005} the diffusion coefficient in the radial direction is reduced by approximately $6$ orders of magnitude with respect to the parallel diffusion coefficient.  The characteristic time estimated for perpendicular diffusion for typical PIII reactor geometries is at the order of a hundred second, while the dissociative recombination characteristic time is around $0.1$ s. Following a similar procedure, one can easily show that mutual neutralization is also likely to dominate diffusion losses provided the negative ion density is of the same order of magnitude as electron density, i.e. moderately electronegative plasmas.
It appears therefore that under PIII discharge conditions, positive ions losses at the reactor wall are dominated by their dissociative recombination with electrons and by their neutralization by collisions with negative ions. 
The investigation of dissociative recombination of molecular ions is therefore of major interest for these processes. This is especially the case of BF$_2^+$,  which is often the major ion in BF$_3$ containing plasma in discharge conditions corresponding to PIII process \cite{young2016}. This study is a continuation of a previous work performed in the theoretical framework of the multichannel quantum defect theory on the dissociative recombination and competitive processes of BF$^+$~\cite{mezei2016}.

The article is organized as follows. After the Introduction, Section II describes
the theoretical approach used in the present calculation. In Section \ref{sec:Cross_sections_rate_coefficients}, the obtained cross sections and the corresponding rate coefficients are displayed and discussed. Section \ref{sec:conclusions_discussions} concludes the study.

\section{Theoretical approach}
\label{sec:theoretical_approach}
\subsection{Dissociative recombination and vibrational excitation cross section formulas} 
  
Since the basic formalism used in our model is presented in detail in Refs. \cite{douguet11a,samantha14}, we restrict ourselves here to underline its major ideas. The studied molecular cation, BF$^+_2$, is linear in its equilibrium geometry.
 
\begin{table*}
\caption{Vibrational frequencies (in cm$^{-1}$) obtained in this study and compared with previous data available in literature.}
\small 
\centering
\begin{tabular}{cccccc}
\toprule
Mode & symmetric stretching & bending & asymmetric stretching & & \\
Symmetry & $\Sigma_{g}^+$ & $\Pi_{u}^+$ and $\Pi_{u}^-$&  $\Sigma_{u}^+$ & &\\
& $\omega_1$	&  $\omega_2$ & $\omega_3$ & Method & Reference \\
\hline
& 1062.8 &  469.2 & 2146.1& CI/ cc-pVTZ & This work \\
& 1023   &  443   & 2088  &  MP2 &   Ref.~\cite{pyykko90}~ \\
& 1030 & 450 &  & CI & Ref.~\cite{peric93}~ \\
&      &     &  2026.1$\pm$ 0.2 & Exp. &  Ref.~\cite{marilyn95}~ \\
\hline
\end{tabular}
\label{tab:frequencies}
\end{table*}

The theoretical model starts with the following assumptions (see for example \cite{samantha14}): (i) the rotation of the molecule is neglected, (ii) the cross section is averaged over the autoionizing resonances, (iii) the autoinization lifetime is assumed to be much longer than the predissociation lifetime ,  and (iv) the harmonic approximation is used to describe the vibrational state of the core ion.
Using (i)-(iv) and applying the frame transformation, the DR cross section is given by Eq.(13) of \cite{samantha14}, in which the scattering matrix elements were expanded to first order in the normal coordinates. 
The cross section for vibrational excitation (VE) of the mode $i$ writes
\begin{eqnarray}
\label{eq:cs_VE}
\sigma^{VE}_i(E_{el})=\frac{\pi\hbar^2}{4 m E_{el}}g_i\sum_{ll'\lambda\lambda' }\left| \frac{\partial S_{l\lambda,l'\lambda'}}{\partial {q}_i}\right|^2 \theta(E_{el}-\hbar\omega_i)\,.
 \end{eqnarray}
Here $q_i$, $\hbar\omega_i$ and $g_i$ $(i=1-3)$ are respectively the dimensionless coordinate, the energy and the degeneracy of the mode $i$. The degeneracy is two for the bending mode 2 and one for the symmetric and asymmetric stretching modes.  $S_{l\lambda,l'\lambda'}$ is an element of the fixed-nuclei scattering matrix for electron--BF$_2^+$ collisions with the initial channel ($\lambda l$) and the exit channel ($\lambda' l'$), $l$ being the electron angular momentum and $\lambda$ its projections on the molecular axis.
And finally, $m$ is the reduced mass of the electron-ion system, $E_{el}$ the incident energy of the electron, $\theta$ the Heaviside step function.
The present theoretical approach can describe the (de-)excitation process changing only one quanta in each normal mode of the target ion. (De-)excitation cross sections for the  changing two or more quanta in a mode, are much smaller (the propensity rule) and are neglected in this study.  In the present case, the initial state of the ion is the ground vibrational level, so the electron can only be captured into the first excited vibrational state of each normal mode.

The situation is similar if the electron energy is not sufficient to excite the ion and then to leave it. In such a situation, the present model suggests that the excitation probability of ion by the electron is described by the same physics, but instead of leaving the vibrationally excited ion, the electron is captured in a Rydberg resonances attached to that vibrational state, excited by the electron. If the electron is captured by the ion, the system will most likely dissociate, rather then autoionize. Correspondingly, the cross section for dissociative recombination (DR) is then obtained \cite{samantha14} as
\begin{eqnarray}
\label{eq:cs_DR}
\sigma^{DR}(E_{el})=\frac{\pi\hbar^2}{4 m E_{el}}\sum_{i=1}^3 g_i\sum_{ll'\lambda\lambda' }\left| \frac{\partial S_{l\lambda,l'\lambda'}}{\partial {q}_i}\right|^2\theta(\hbar\omega_i-E_{el})\,.
 \end{eqnarray}
Here $i$ runs over all three modes:  two stretching modes $\nu_1$ and $\nu_3$ (symmetric and asymmetric) with respective frequencies $\omega_1$ and $\omega_3$ and corresponding coordinates $q_1$ and $q_3$, and a doubly degenerate transverse mode $\nu_2$ with a lower frequency $\omega_2$ and coordinates $(q_{2x},q_{2y})$.

To calculate the derivative of the scattering matrix $\partial S_{l\lambda,l'\lambda'}/\partial q_i$ with respect to the normal coordinate $q_i$, the scattering matrix is evaluated for two values of $q_i$ keeping the other normal coordinates $q_{i'}$ fixed at $q_{i'}=0$.

The elements  $ S_{l\lambda,l'\lambda'}(\vec q)$ of the scattering matrix $\hat S(\vec q)$ at a given geometry $\vec q$ specified by four normal coordinates $\vec q=\{q_1,q_2,q_3,q_4\}=\{q_1,q_{2x},q_3,q_{2y}\}$ 
are computed from the reactance matrix $\hat K$, obtained numerically as discussed below:
\begin{eqnarray}
\hat S =\frac{\hat 1 + i \hat K}{\hat 1 - i \hat K}\,,
\end{eqnarray}
where $\hat 1$ is the identity matrix.

\subsection{The properties of the BF$_2^+$ ion and the scattering calculations}

The main electronic ground state configuration $^1\Sigma_g^+$ of the ion in its natural point group symmetry $D_{\infty h}$ is $\left(1\sigma_{u}^+\right)^2\left(1\sigma_{g}^+\right)^2 \left(2\sigma_g^+\right)^2 \left(2\sigma_{u}^+\right)^2 \left(3\sigma_{g}^+\right)^2 \left(4\sigma_{g}^+\right)^2 \left(3\sigma_{u}^+\right)^2\left(1\pi_{u}^-\right)^2$ $\left(1\pi_{u}^+\right)^2 \left(1\pi_{g}^-\right)^2 \left(1\pi_{g}^+\right)^2$. The normal coordinates and the related frequencies are obtained using the cc-pVTZ basis set centered on each atom and including $s$, $p$ and $d$ orbitals. Performing configuration interaction (CI) calculation in the $D_{2h}$ symmetry group, using the MOLPRO suite of codes \cite{MOLPRO_brief}, we found equilibrium geometry of the ion with the B-F distance of $1.2215$ \AA. Table \ref{tab:frequencies} gives the vibrational frequencies obtained from CI calculations using cc-pVTZ basis set, and compares with the data available in literature. 

\begin{figure*}
\vspace*{1.5cm}
\includegraphics[width=0.85\textwidth]{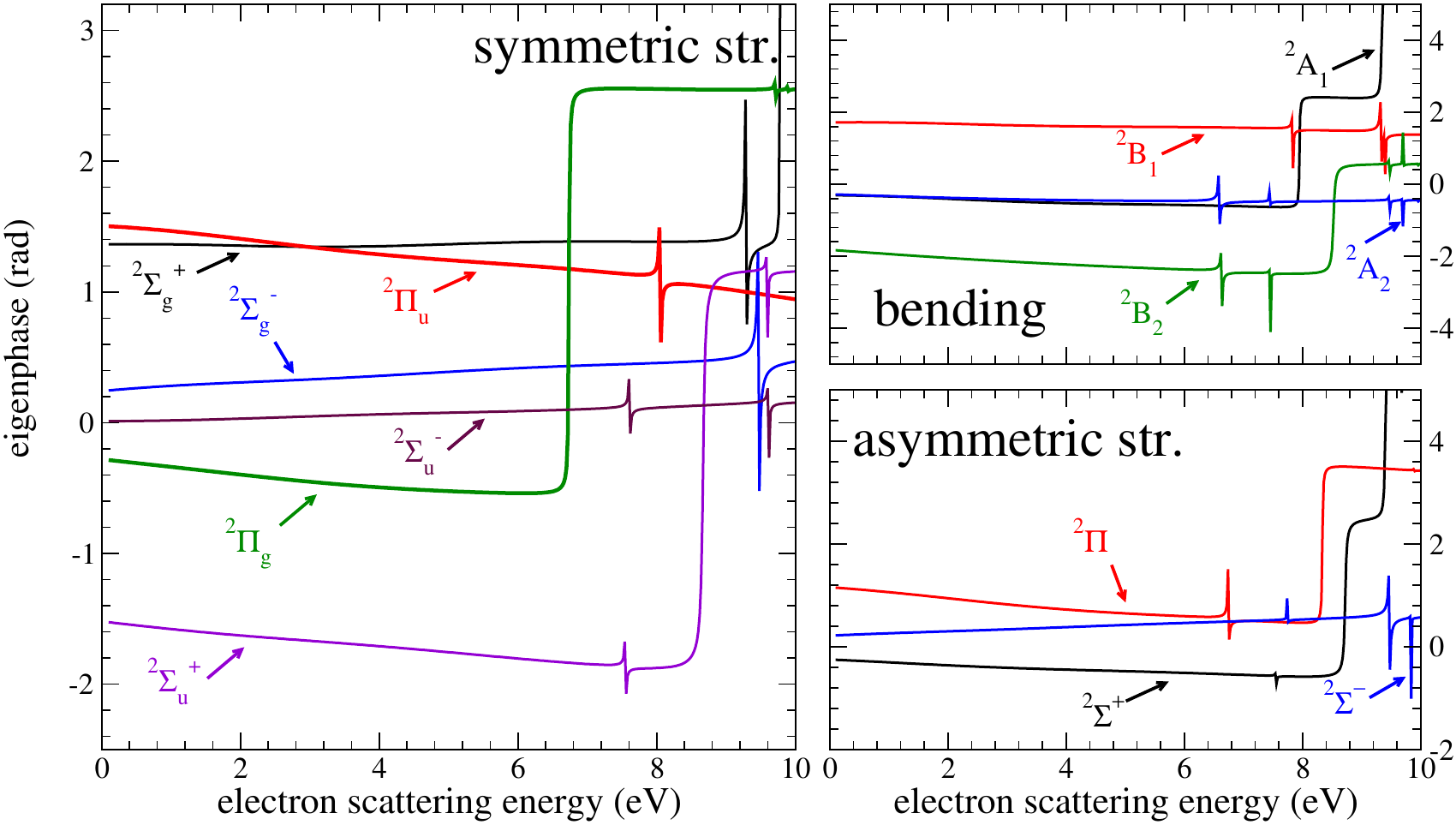}
\vspace*{0.5cm}
\caption{\label{fig:eigenphase} Eigenphase sums as a function of the electron scattering energy $E_{el}$ for $q_i=0.01$ (dimensionless) for the symmetric stretching (left panel), bending (right upper panel), and asymmetric stretching (right bottom panel) modes. The curves of different color correspond to different symmetries of the $e^-$--BF$_2^+$ system. 
}
\vspace{0.5cm}
\end{figure*}

The electron scattering calculations were performed using the UK R-Matrix code \cite{tennyson10,carr12} with the Quantemol-N interface \cite{tennyson07quantemol}. The calculations were performed in the abelian subgroup $D_{2h}$ and the target BF$_2^+$ ion was assumed to be in its ground electronic state. In our CI model, we freeze 10 electrons in the core $1a_g, 2a_g, 3a_g, 1b_{2u}, 2b_{2u}$, while the remaining 12 electrons are kept free in the active space of $4a_g, 5a_g, 1b_{3u}, 2b_{3u}, 3b_{2u}, 4b_{2u}, 1b_{1g}, 1b_{1u}, 2b_{1u}$, $1b_{3g}$  molecular orbitals. A total of eight electronic excited target states are represented by 1844 configuration state functions (CSFs) for the ground state. All the generated states up to 16 eV were retained in the final close-coupling calculation. We used an R-matrix sphere of radius 10 bohrs and continuum Gaussian-type orbitals with partial waves up to $l<4$. In the following, this calculation will be referred to CAS$_1$. 

 Initially, several basis sets - including DZP and cc-pVTZ - were tested to investigate the stability of the target properties such as polarizability and ground state energy and, finally, we chose the cc-pVTZ basis set in order to perform the scattering calculations. Since the first electronically excited state $^3 \Sigma_{g}^-$ is approximately 13 eV above the dissociation limit for the ground state, this latter state is essentially isolated and the non-adiabatic effects are expected to be small. Therefore, for low electron energy collisions,  i.e. bellow 10 eV, only the ground
electronic state is open for ionization in e--BF$_2^+$ collisions, and the dimension of the geometry-fixed scattering matrix remains the same at low collision energies.  

At low collision energies the fixed-nuclei scattering matrix depends only weakly on energy. A  sharper energy-dependence is observed at certain relatively-high energies, corresponding to positions of Rydberg states attached to the excited electronic states of the ion. A  quantity convenient for the analysis of the energy dependence of the scattering matrix is the eigenphase sum. Figure~\ref{fig:eigenphase} displays the eigenphase sum for the three different geometries corresponding to a small displacement from equilibrium along each normal mode of the BF$_2^+$ ion. Bending and asymmetric stretching mode calculations were performed in the $C_{2v}$ abelian subgroup, while for the symmetric mode the group $D_{2h}$ was used in the calculations. The variation of the eigenphase sums is smooth for energies below 6 eV. Above this value, a sharp energy dependence at certain energies is observed  due to the presence of electronic Rydberg resonances attached to closed ionization limits. 

\section{Cross sections and rate coefficients}
\label{sec:Cross_sections_rate_coefficients}
  
\begin{figure}
\centering
\includegraphics[width=0.95\columnwidth]{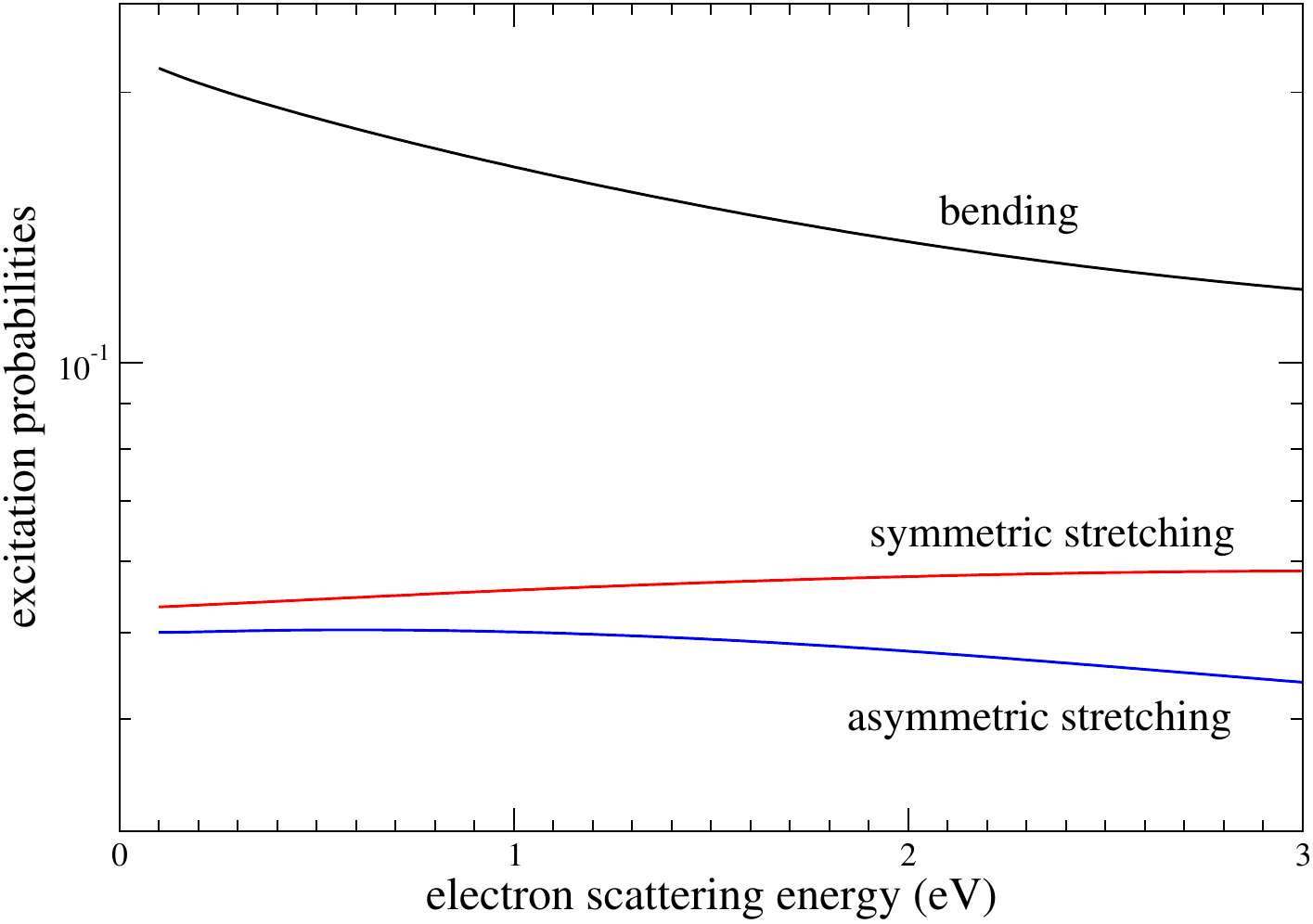}
\caption{\label{fig:probabilities} Vibrational excitation of BF$_2^+$: probabilities corresponding to the normal vibrational modes of the target ion.}
\end{figure}

For convenience, we introduce the quantities
\begin{equation}
 P_i=\frac{g_i}{2}\sum_{ll'\lambda\lambda' }\left| \frac{\partial S_{l\lambda,l'\lambda'}}{\partial {q}_i}\right|^2\,,
\label{eq:Pi}
\end{equation}
which could be interpreted as the probability for excitation of the vibrational mode $i$, are listed in table \ref{tab:rate_coefficients_parameters}. Figure \ref{fig:probabilities} shows the weakly dependence of that quantity on energy and, therefore, could be used as a constant in calculations of the thermally-averaged rate coefficients. 

\begin{table}
\caption{Parameters of Eqs.~(\ref{eq:rates}) and (\ref{eq:rates2}) calculated at $E=0.1$ eV collision energy.} 
\small 
\centering
\begin{tabular}{cc}
\toprule
Mode $i$ &  $P_i$  \\
\hline
symmetric stretching &	  0.053  \\
bending &  0.20  \\
asymmetric stretching &	   0.05\\
\hline
\end{tabular}
\label{tab:rate_coefficients_parameters}
\end{table}

Using  $P_i$, the cross sections of Eqs.~(\ref{eq:cs_VE}) and (\ref{eq:cs_DR}) are written as   
\begin{eqnarray}
\label{eq:cs_VE2}
\sigma^{VE}_i(E_{el})=\frac{\pi\hbar^2}{2 m E_{el}}P_i\ \theta(E_{el}-\hbar\omega_i)\,, \\
\sigma^{DR}(E_{el})=\frac{\pi\hbar^2}{2 m E_{el}}\sum_{i=1}^3P_i\ \theta(\hbar\omega_i-E_{el})\,.
 \end{eqnarray}

\begin{figure}
\centering
\includegraphics[width=0.95\columnwidth]{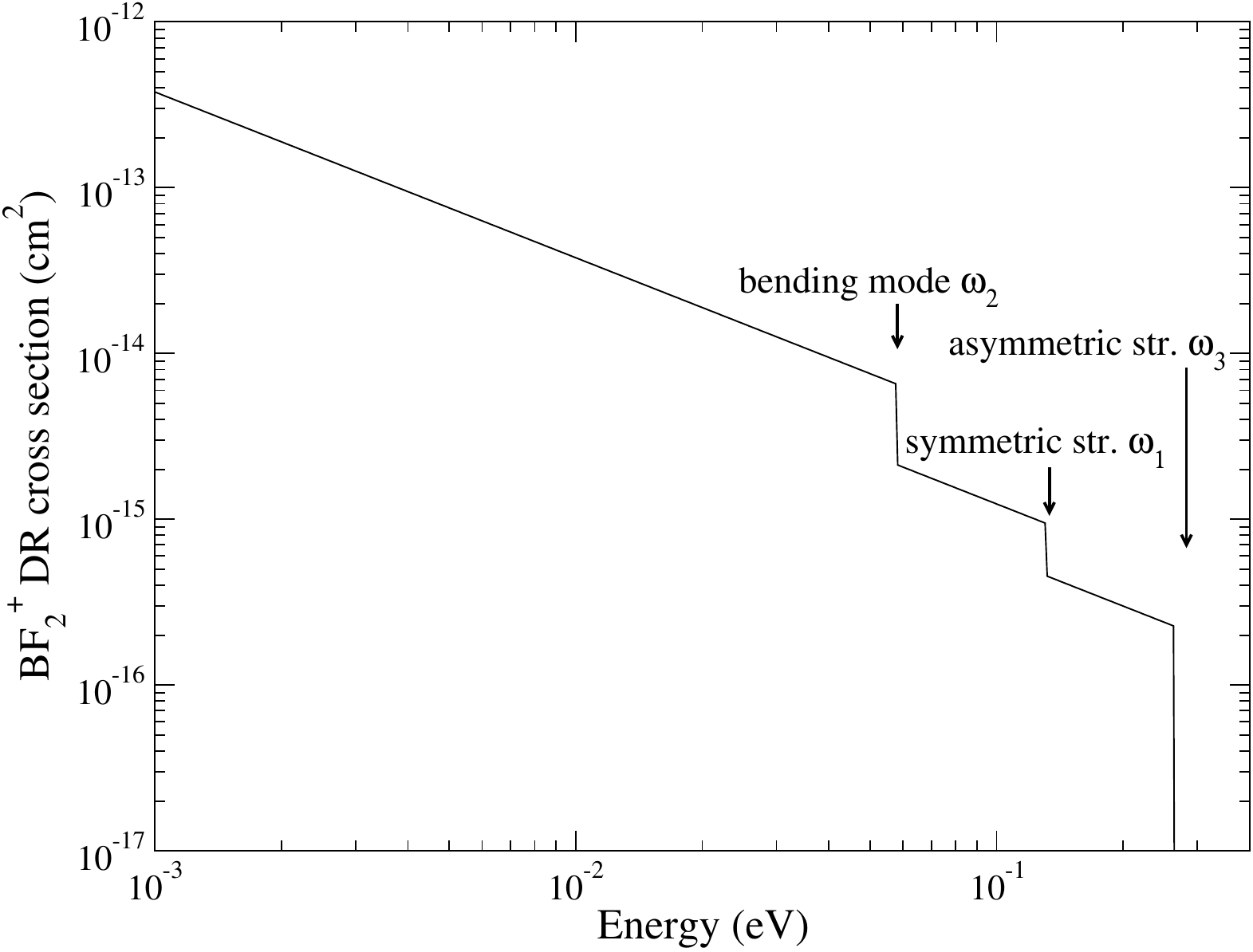}
\caption{\label{fig:xs_DR_BF2p} Cross section for the dissociative recombination of BF$_2^+$. The lowest vibrational threshold of each normal mode are indicated by arrows. }
\end{figure}

Figure \ref{fig:xs_DR_BF2p} illustrates the DR cross section $\sigma^{DR}(E_{el})$  computed based on Eq. (\ref{eq:cs_VE2}). At very low scattering energies, i.e. below 0.02 eV, the DR cross section is a smooth function inversely proportional to the incident energy of the electron, as predicted by the Wigner threshold law, whereas at higher energies, it exhibits a drop at each vibrational threshold. 
 
\begin{figure}
\centering
\includegraphics[width=0.95\columnwidth]{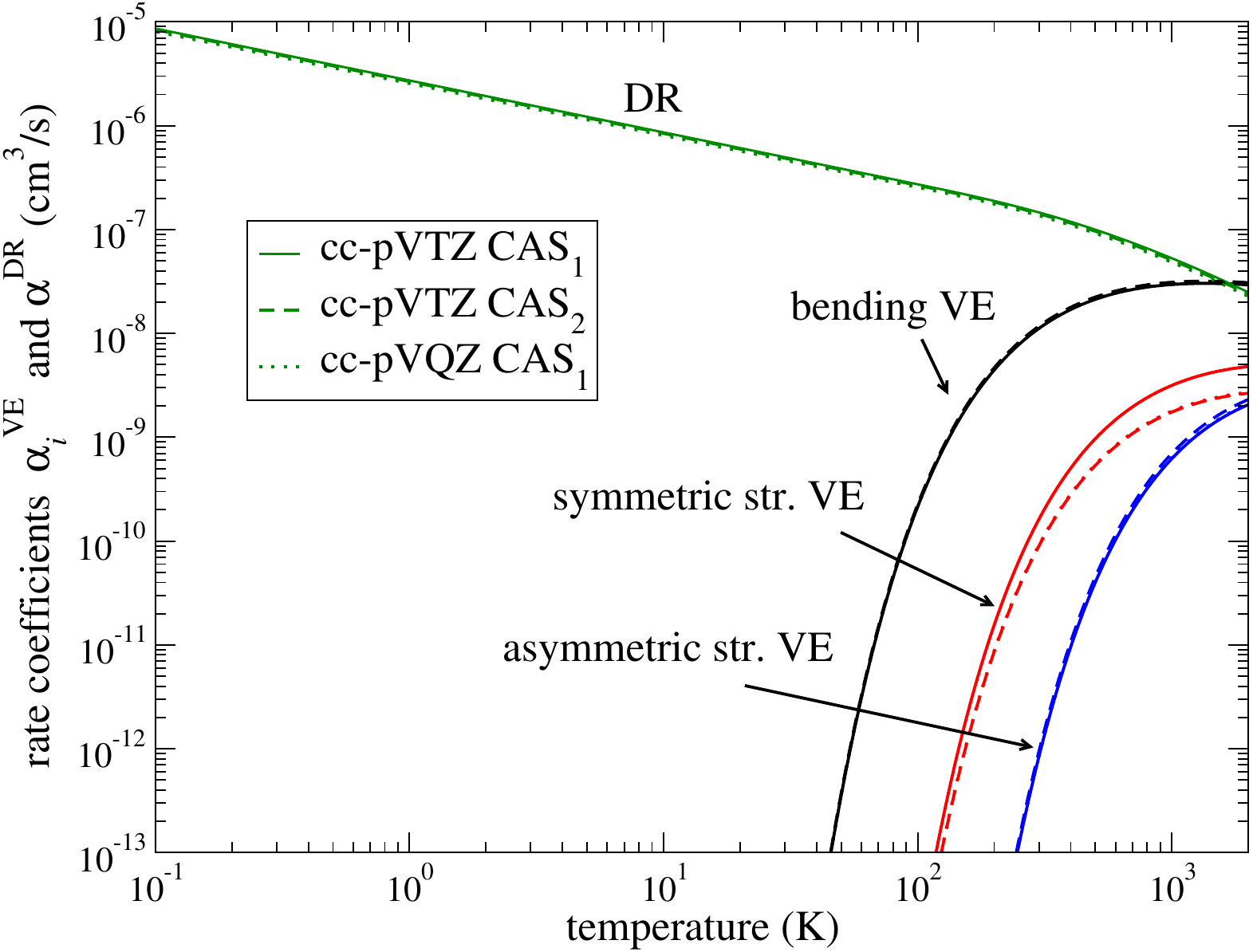}
\caption{\label{fig:rates} Dissociative recombination and vibrational excitation of BF$_2^+$: rate coefficients - Eqs.~(\ref{eq:rates}) and (\ref{eq:rates2}). To give an idea about the uncertainty of the present results, we also plotted results of two calculations with the cc-pVTZ CAS$_2$ (dashed line) and the cc-pVQZ CAS$_1$ (dotted line) sets of parameters of the model. For the vibrational excitation of the symmetric stretching mode, the cc-pVTZ CAS$_2$ and cc-pVQZ CAS$_1$ curves are indistinguishable and slightly below the cc-pVTZ CAS$_1$ curve.  For the remaining three processes, the three calculations produce the curves almost indistinguishable in the figure.}
\end{figure}

Due to the simple analytical form of the cross sections, the corresponding rate coefficients are easily evaluated from the general expression 
\begin{eqnarray}
\label{eq:rate}
\alpha(T)=\frac{8\pi }{\left(2\pi k_b T\right)^{3/2}} \int_0^\infty \sigma(E_{el}) \exp\left(-\frac{E_{el}}{k_b T}\right) E_{el} dE_{el}\,,
 \end{eqnarray}
giving
\begin{eqnarray}
\label{eq:rates}
\alpha^{VE}_i(T)=\sqrt{\frac{2\pi}{k_b T}}\frac{\hbar^2}{m^{3/2}}\ P_i\ \exp\left(-\frac{\hbar\omega_i}{k_b T}\right)\,,\\
\alpha^{DR}(T)=\sqrt{\frac{2\pi}{k_b T}}\frac{\hbar^2}{m^{3/2}}\ \sum_{i=1}^3 P_i\ \left[1-\exp\left(-\frac{\hbar\omega_i}{k_b T}\right)\right]\,.\label{eq:rates2}
 \end{eqnarray}
where $k_b$ is the Boltzmann coefficient and $T$ is the temperature. The thermally averaged rate coefficients for dissociative recombination and vibrational excitation are shown in Figure \ref{fig:rates}. 

In order to access the uncertainty of the present theoretical model, we have performed a complete calculation of the DR and VE rate coefficients using different basis sets and orbital space in the electron-scattering calculations. The calculations were performed for three sets of parameters (1) the CAS$_1$ with the cc-pVTZ basis set, mentioned above; (2) a calculation (referred here as cc-pVTZ CAS$_2$) similar to (1) but with a smaller orbital space, where 8 electrons are kept free in the active space; and (3) a calculation (referred here as cc-pVQZ CAS$_1$) similar to (1) but with the larger basis cc-pVQZ. The results are in shown in Fig.~\ref{fig:rates}. The difference between the rate coefficients produced in the three calculations for the DR process and the VE of the asymmetric stretching and bending modes is about $4\%$. The uncertainty of the rate coefficient for the VE of the symmetric stretching mode is larger, varying in the interval $10-40\%$ for different temperatures. Notice that the overall probability for the symmetric stretching excitation is much smaller than the probabilities for other modes and DR.

\begin{figure}
\centering
\includegraphics[width=0.95\columnwidth]{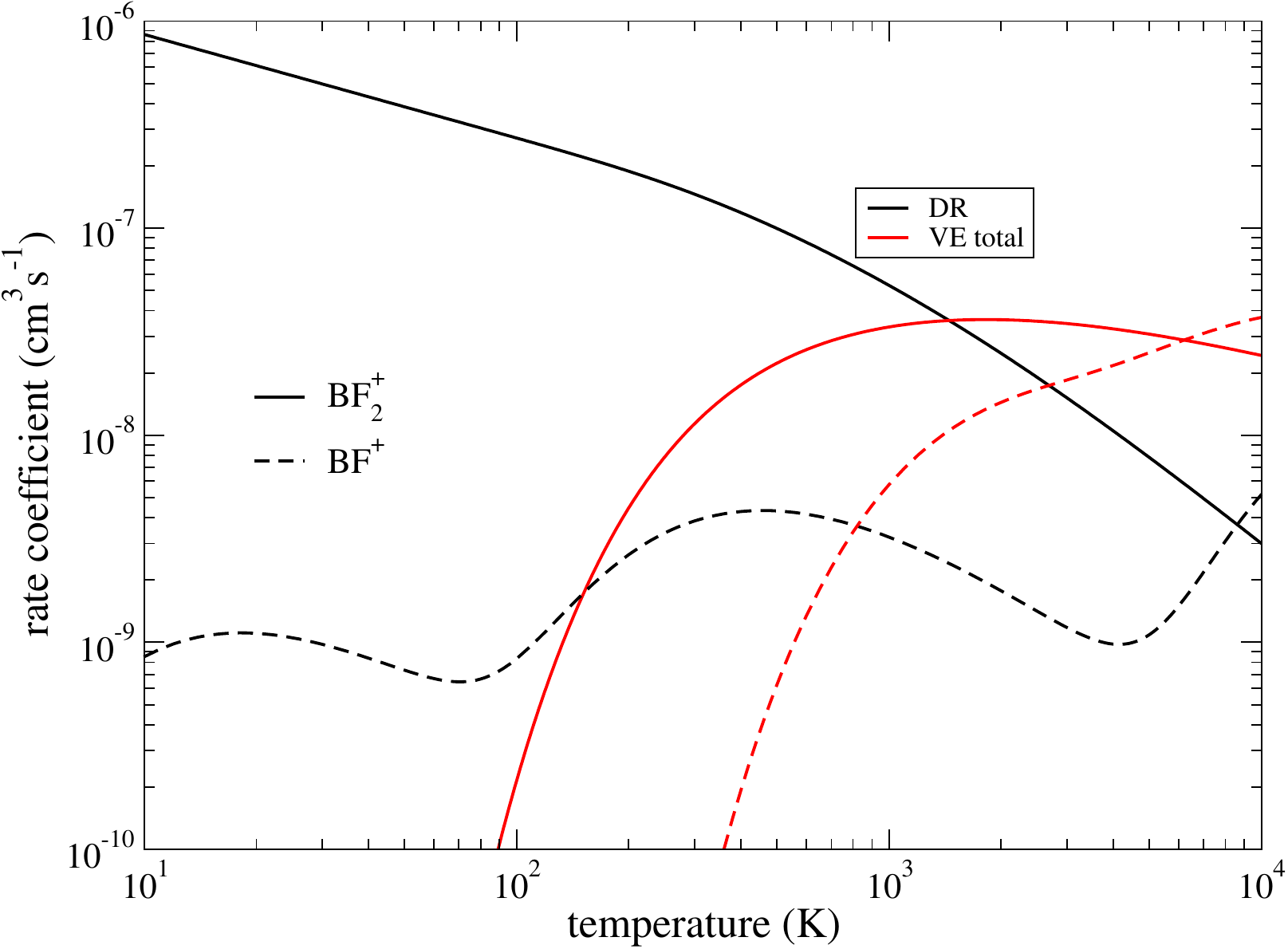}
\caption{\label{fig:relative_importance} Relative importance of the dissociative recombination and vibrational excitation of BF$_2^+$, with respect to those of  BF$^+$.}

\end{figure}

\section{Conclusions and discussions}
\label{sec:conclusions_discussions}

In this study, cross sections and rate coefficients for dissociative recombination and vibrational excitation of BF$_2^+$ by electron-impact were obtained using a theoretical approach that combines the normal modes approximation for the vibrational states of the target ion, the vibrational frame transformation, and the UK R-matrix code. The thermally averaged rate coefficients have a simple analytical form.  

The obtained thermally-averaged rate coefficients are relevant for the kinetic modeling of molecule based cold non-equilibrium plasmas, in the context of complete lack of other theoretical or experimental data on these processes for this cation, and are ready to be used in the modeling of the fluorine/boron plasma for etching or implantation processes. Indeed, we are presently able to make important statements on the relative importance of BF$_2^+$ with respect to BF$^+$ on the population and excitation balance. In particular, as shown in Figure 5, the dissociative recombination of BF$_2^+$ strongly dominates that of BF$^+$ below 7000 K, and the vibrational excitation displays the same feature below 5000 K.  

The rotational structure of the target ion and of the neutral molecule was neglected in the present approach, which implies that the obtained cross sections and rate coefficients should be viewed as averaged over initial rotational states and summed over final rotational states of the corresponding initial and final vibrational levels (for vibrational excitation) or  dissociative states (for dissociative recombination). Purely rotational transitions, i.e. without changing the vibrational state, might be useful to model very cold environment, below 40 K, which is not the case for the presently investigated BF$_3$ plasma. BF$_2^+$ has no permanent dipole, so the rotational transitions are likely to have very small cross sections.

\section*{Acknowledgments}
This work was supported by the National Science Foundation, Grant No PHY-18-06915. 
Ioan F. Schneider thanks for generous financial support from La R\'egion Haute-Normandie, via  the GRR Electronique, Energie et Mat\'eriaux and the project BIOENGINE, from the F\'ed\'eration de Recherche ''Energie, Propulsion, Environnement", 
and from the LabEx EMC$^3$ and FEDER \textit{via} the projects PicoLIBS (ANR-10-LABEX-09-01), EMoPlaF and CO$_2$-VIRIDIS. 
Ioan F. Schneider and  J\'anos Zsolt Mezei acknowledge support from 	the CNRS \textit{via} the GdR THEMS, IAEA (Vienna) \textit{via} the Coordinated Research Project ''Light Element Atom, Molecule and Radical Behaviour in the Divertor and Edge Plasma Regions", the Programme National ''Physique et Chimie du Milieu Interstellaire'' (PCMI) of CNRS/INSU with INC/INP co-funded by CEA and CNES, and F\'ed\'eration de Recherche ''Fusion par Confinement Magn\'etique" (CNRS and CEA). J\'anos Zsolt Mezei and Khalid Hassouni acknowledge support from USPC \textit{via} ENUMPP and Labex SEAM. Ioan F. Schneider thanks Laboratoire  Aim\'e Cotton for outstanding hospitality. 

\section*{Data availability}
Upon a reasonable request, the data supporting this article will be provided by the corresponding author.

\bibliography{DR}

\end{document}